\documentclass[9pt,conference]{IEEEtran}
\IEEEoverridecommandlockouts
\usepackage{cite}
\usepackage{amsmath,amssymb,amsfonts}
\usepackage{algorithmic}
\usepackage{graphicx}
\usepackage{textcomp}
\usepackage{xcolor}
\usepackage{svg}
\usepackage{tipa}
\usepackage{colortbl}
\usepackage{multirow,arydshln}
\usepackage{tablefootnote}
\usepackage{enumitem}
\usepackage{booktabs} 
\usepackage{verbatim}
\usepackage{hyperref}
\usepackage{balance}

\usepackage{soul}

\def\BibTeX{{\rm B\kern-.05em{\sc i\kern-.025em b}\kern-.08em
    T\kern-.1667em\lower.7ex\hbox{E}\kern-.125emX}}
\begin{document}

\title{Exploring Acoustic Similarity in Emotional Speech and Music via Self-Supervised Representations}

\makeatletter
\newcommand{\linebreakand}{%
  \end{@IEEEauthorhalign}
  \hfill\mbox{}\par
  \mbox{}\hfill\begin{@IEEEauthorhalign}
}
\makeatother

\author{
\IEEEauthorblockN{Yujia Sun, Zeyu Zhao, Korin Richmond, Yuanchao Li$^\dag$\thanks{$^\dag$Corresponding author. \textit{yuanchao.li@ed.ac.uk}}}
\IEEEauthorblockA{\textit{University of Edinburgh, UK}}
}

\maketitle

\begin{abstract}
Emotion recognition from speech and music shares similarities due to their acoustic overlap, which has led to interest in transferring knowledge between these domains. However, the shared acoustic cues between speech and music, particularly those encoded by Self-Supervised Learning (SSL) models, remain largely unexplored, given the fact that SSL models for speech and music have rarely been applied in cross-domain research. In this work, we revisit the acoustic similarity between emotion speech and music, starting with an analysis of the layerwise behavior of SSL models for Speech Emotion Recognition (SER) and Music Emotion Recognition (MER). Furthermore, we perform cross-domain adaptation by comparing several approaches in a two-stage fine-tuning process, examining effective ways to utilize music for SER and speech for MER. Lastly, we explore the acoustic similarities between emotional speech and music using Fréchet audio distance for individual emotions, uncovering the issue of emotion bias in both speech and music SSL models. Our findings reveal that while speech and music SSL models do capture shared acoustic features, their behaviors can vary depending on different emotions due to their training strategies and domain-specificities. Additionally, parameter-efficient fine-tuning can enhance SER and MER performance by leveraging knowledge from each other. This study provides new insights into the acoustic similarity between emotional speech and music, and highlights the potential for cross-domain generalization to improve SER and MER systems.

\end{abstract}


\section{Introduction}
\label{sec:intro}

Emotion recognition from audio signals has become a pivotal task in various applications, ranging from human-computer interaction to affective computing. Two important subfields in this domain are Speech Emotion Recognition (SER) and Music Emotion Recognition (MER). While SER focuses on identifying emotional states from speech, MER aims to recognize emotions conveyed through music. Both tasks share common challenges, such as dealing with subjective emotional labels and the variability of acoustic signals, yet they offer complementary insights into how emotions are expressed and perceived across different modalities \cite{cowie2001emotion,scherer2000cross,kim2010music,yang2012machine}.

Furthermore, there has been recognition of the synergy between SER and MER. Both tasks involve the analysis of temporal, spectral, and emotional cues in audio, and advances in one field can benefit the other. For instance, insights gained from modeling emotional dynamics in speech can inform MER, particularly in understanding how rhythm, pitch, and timbre contribute to emotion perception \cite{coutinho2017shared}. By leveraging shared architectures and pretraining strategies, it has led to the feasibility of transfer learning and joint training  for SER and MER with cross-domain knowledge \cite{zhang2015recognizing,zhang2016cross}.

Recently, Self-Supervised Learning (SSL) has emerged as a powerful paradigm in audio analysis, addressing the need for large annotated datasets by leveraging unlabeled data \cite{baevski2022data2vec,zhu2021musicbert}. Self-supervised models have demonstrated remarkable success by pretraining on vast amounts of unlabeled audio data, enabling the extraction of effective representations for downstream tasks, including SER and MER. These models have significantly advanced SER by capturing nuanced prosodic and paralinguistic features \cite{li2023exploration}, and are now increasingly being applied to MER, where they show promise in learning emotional patterns from music \cite{li2023mert}.

Nevertheless, the shared acoustic cues between speech and music, particularly those encoded by SSL models, remain largely unexplored. While SSL-based models have demonstrated impressive performance in both tasks, the extent to which these models capture and utilize overlapping acoustic features across speech and music is not well understood. This gap limits progress in developing cross-domain insights and hinders the potential for improving SER and MER systems.

Therefore, we address two main research questions: \textit{\textbf{1)} What insights into SSL models can be gained from their performance in SER and MER?} \textit{\textbf{2)} Can cross-domain generalization improve performance in SER and MER through domain adaptation?} To answer these questions, we undertake the following tasks in both the speech and music domains:

$\bullet$ We conduct cross-domain layerwise probing utilizing SSL models pretrained on either speech or music data.

$\bullet$ We implement domain adaptation techniques in a two-stage process to improve emotion recognition performance with limited data for each task.

$\bullet$ We evaluate the acoustic similarity between speech and music for each emotion using SSL representations with Fréchet audio distance.

% Note that we focus on song, a subdomain of music, as we aim to reduce the impact of instrumentation for a fair evaluation of the acoustic similarity.

Note that in this work we focus on song, a subdomain of music, since the dataset we use contains vocal-only music. This reduces the impact of instrumentation and supports a fair evaluation of the similarity between speech and music in terms of vocal acoustics.

% a layer-wise probing experiment and a domain adaptation experiment. We utilize speech and song data from the RAVDESS dataset \cite{livingstone2018ravdess}, employing the SSL models Wav2Vec 2.0 \cite{Baevski2020wav2vec2} and HuBERT \cite{Wei2021hubert} for speech, and MERT \cite{li2024mertacousticmusicunderstanding} for music. Additionally, we compare the audio similarity between speech and song in various emotional contexts by calculating cross-domain Fréchet Audio Distance (FAD) scores \cite{kilgour2019frechetaudiodistancemetric}.

% In this context, we investigate whether this speech-music relatedness can enhance machine performance in SER and MER. In our work, we experiment with a domain adaptation strategy inspired by \cite{Lashkarashvili_2024}.

% The structure of this paper is as follows: Section 2 reviews previous analyses of SSL models and explores the relationship between speech and song. Section 3 details the dataset and experimental design. Sections 4 and 5 present the results of the layer-wise probing and domain adaptation experiments, respectively. Section 6 discusses the cross-domain FAD score comparisons across emotions. Finally, Section 7 concludes with our findings.

\section{Related Work}
Over the past decade, researchers have explored the acoustic similarities and generalizability between speech and music. \cite{biqiao2015predictionemo} demonstrated that models leveraging cross-domain features outperformed those relying on domain-specific features, suggesting that speech and music share certain emotional characteristics. Subsequently, they applied multitask learning to jointly predict emotions from both speech and music \cite{zhang2015recognizing,zhang2016cross}. In another study, \cite{coutinho2017shared} investigated transfer learning between the two domains using denoising autoencoder-pretrained long short-term memory networks for arousal-valence regression tasks, showing promising cross-domain generalizability. Likewise, \cite{gomez2020emotionrecognition} found that pretraining on speech data enhanced the performance of convolutional neural networks in MER.

Nevertheless, the exploration of acoustic similarity between speech and music, as well as the use of transfer learning and domain adaptation between SER and MER, remains limited, especially in the era of SSL models. Although SSL models have demonstrated success in various audio-related tasks, their application to such cross-domain research has yet to be explored. Therefore, we revisit the utility of acoustic correlations between emotional speech and music by using representations from SSL models pretrained on either speech or music.

% \subsection{Layerwise Analysis}
% To tackle the opacity of deep learning machines, many have investigated their layerwise ability. 
% % \cite{Pasad2021layerwise, Pasad2023comparative} found that shallow layers of \ac{W2V2} retain acoustic information from the input signal, intermedia layers contain segmental cues, later layers semantics, and final layers back to acoustics, while in HuBERT, segmental and semantic information appears to concentrate in more later layers. 
% For prosody-related tasks, \cite{Li2023emotion} found that representations output by middle layers from W2V2 perform the best in SER compared to other layers, while \cite{lin2023prosody} found the contribution of latter layers are most salient for HuBERT. Here, we continue to explore SSL models' layerwise behaviour through linear probing. Extending the analysis to MERT also allows us to compare the effects of different pretraining data types. Since \ac{W2V2}, HuBERT, and MERT all have very similar architectures, our results reflect the influence of their pretraining objectives on their behaviour \cite{Chung2021similarity}.

% \footnote{Among these studies, \cite{biqiao2015predictionemo} used vocal-only datasets, while \cite{Coutinho2017sharedacoutic, gomez2020emotionrecognition} used music datasets of songs with instrumental music.}

\section{Dataset and SSL Models}
\subsection{Dataset}
We use the RAVDESS dataset \cite{livingstone2018ravdess}, an audio-visual collection consisting of acted affective recordings of speech and song. The dataset features recordings of 12 female and 12 male professional actors speaking English with a North American accent. In our study, we focus on the audio recordings, excluding the visual input. These recordings contain only vocal performances without accompanying instrumental music, thereby excluding the instrumental influence on the acoustic similarity. Since the song recordings of one female actor are missing, we discard her speech recordings as well to ensure an equal amount of speech and song data for model training. The text content in both the speech and song recordings is semantically neutral and identical, allowing us to concentrate on the acoustic properties by eliminating potential effects caused by lexical content (e.g., prosody variations due to word pronunciation). These are the key reasons we have chosen RAVDESS, despite the availability of larger datasets.

For emotion categories, we select the ones common to both the speech and music recordings, including \textit{neutral}, \textit{calm}, \textit{happy}, \textit{sad}, \textit{angry}, and \textit{fearful}. This yields 92 files for neutral and 184 files for each of the other emotions in both domains. The higher count for emotions other than neutral is due to their having two intensity levels, whereas neutral is presented with only one normal intensity. To ensure sufficient data samples, we ignore the intensity distinction and randomly select 60\% of the data for training, 20\% for validation, and 20\% for testing, from the total of 1,012 recordings in each domain.

\subsection{SSL Models}
We select \href{https://huggingface.co/facebook/wav2vec2-base-960h}{\textit{Wav2Vec 2.0-Base-960h}} (W2V2), \href{https://huggingface.co/facebook/hubert-base-ls960}{\textit{HuBERT-Base-ls960}} (HuBERT) \cite{hsu2021hubert}, and \href{https://huggingface.co/m-a-p/MERT-v1-95M}{\textit{MERT-v1-95M}} (MERT) for investigation. All three models have a very similar architecture, consisting of seven convolutional encoders and 12 transformer encoders with around 95M parameters in total. Hidden representations from all 12 layers are used as the input to the downstream classifier for SER and MER.

The primary difference among the three models lies in their pretraining objectives. W2V2 masks the speech input in the latent space and solves a contrastive task defined over a quantization of the latent representations, which are jointly learned. HuBERT shares the same concept as W2V2 but applies a classification task, requiring the model to classify hidden sequences into predefined K-means clusters. Both W2V2 and HuBERT are pretrained on LibriSpeech \cite{kahn2020libri}, whereas MERT is pretrained on 160k hours of music recordings from the Internet. Although MERT’s pretraining strategy generally aligns with that of HuBERT, its objective is modified to incorporate music-relevant cues, such as harmony, timbre, and musical pitch.

% The primary difference among the three models is in their pretraining objectives. W2V2 employed \ac{CPC}
% % : it learns a distribution of features so that, when given a masked position in a sequence, it can identify the correct representation at this step from a few distractors sampled from other positions of the same sequence \parencite{Baevski2020wav2vec2}
% . HuBERT employed \ac{MPC}
% % : it learns how to minimise the MFCC-based difference between its prediction at a masked position and the actual signal (i.e., learn to recover the masked representation) \parencite{Wei2021hubert}
% . The speech datasets used to pretrain both W2V2 and HuBERT in a self-supervised manner are LibriSpeech \cite{libri2020} and Libri-light \cite{librilight2020}. Contrastively, the dataset for pretraining MERT is Music4All \cite{music4all2020} consisting of songs with instrumental music. MERT's pretraining method generally follows that of HuBERT, yet \cite{li2023mert} made the objective consider more music-relevant cues such harmonic, timbre, and musical pitch.

\section{Methodology}
\label{sec:pagestyle}

\subsection{Layerwise Probing of Emotion Recognition Performance}
To investigate the layerwise behavior of each model on either speech or music data, we first extract features from each of the three upstream SSL models (W2V2, HuBERT, or MERT) and apply mean pooling over all frames along the time dimension to these features, following previous work by \cite{li2023exploration,saliba2024layer}. Subsequently, for each layer, we train a linear classifier on the training set to perform SER or MER. We select the best-performing checkpoint based on the Unweighted Accuracy (UA) on the validation set and evaluate it on the test set.

\subsection{Domain Adaptation for Cross-Domain Performance Improvement}
To investigate whether shared acoustic cues between speech and music can be transferred for emotion recognition in each domain, we combine all 12-layer representations and implement three approaches for transferering the shared acoustics:

$\bullet$ \textbf{\textit{1)}} \textbf{Baseline}: we keep the pretrained SSL model frozen and perform mean pooling across the 12-layer representations. Only the linear classifier is trainable.

$\bullet$ \textbf{\textit{2)}} \textbf{Weighted-Sum} (WS): we replace mean pooling with learnable 12-dimensional WS parameters, while the SSL models maintain frozen, consistent with the baseline.
% \YC{Is the SSL models frozen?}

$\bullet$ \textbf{\textit{3)}} \textbf{Parameter-Efficient Fine-Tuning} (PEFT): we incorporate WS, Weighting-Gate (WG), Low-Rank Adapter (LoRA) \cite{hulora}, and Bottleneck Adapter (BA) \cite{houlsby_parameter-efficient_2019}. The SSL models are kept frozen, with only the PEFT modules being trainable.

% \begin{figure}[htb]

% \begin{minipage}[b]{1.0\linewidth}
%   \centering
%   \centerline{\includesvg[width=8.5cm, inkscapelatex=false]{figures/transformer.svg}}
% \end{minipage}
% \caption{Implementation of PEFT adapters following \cite{Lashkarashvili_2024}. Hidden states$_n$ is the output from the $n$th transformer layer, or that from the preceding projection layer when $n=0$. We only use $n$ from one to 12 for downstream tasks. We use $r=8$ for LoRA and set the bottleneck dimension to 48.}
% \label{fig:peft}
% %
% \end{figure}

The same downstream classifier used in layerwise probing is employed in all approaches. As two-stage fine-tuning has demonstrated cross-corpus and cross-lingual SER ability \cite{lashkarashvili2024parameter,han2024cross}, we apply this approach to all three methods. Specifically, in \textit{Stage One}, we train the model on the source domain (either speech or music), save the model with the best validation accuracy, and test it on the source domain. In \textit{Stage Two}, we load the saved model, further train it on the target domain (music if the source is speech, or vice versa), and again save the model with the highest validation accuracy, testing it on the target domain.

\subsection{Layerwise Analysis of Cross-Domain Acoustic Similarity}
To explore the extent to which acoustic representations are shared between emotional speech and music, we adopt Fréchet Audio Distance (FAD) as a reference-free measure to assess acoustic similarity. Compared to traditional similarity metrics, such as cosine similarity and Euclidean distance, FAD is specifically designed for audio assessment, capturing the perceptual similarity between two audio embedding distributions \cite{roblek2019fr}. It has been shown to effectively distinguish between real and synthetic audio \cite{gui2024adapting}, as well as audio with different emotions \cite{li2024rethinking}. Therefore, we use FAD to evaluate the acoustic similarity between speech and song. The calculation is as follows.

Given the embeddings of speech set and music set, $\mathbf{X}^{s}$ and $\mathbf{X}^{m}$, the FAD score is calculated using multivariate Gaussians derived from the two embedding sets $X^{s}(\mu_s, \Sigma_s)$ and $X^{m}(\mu_m, \Sigma_m)$:
\begin{align}
F(X^{s}, X^{m}) = ||\mu_s - \mu_m|| ^2 + tr(\Sigma_s + \Sigma_m - 2\sqrt{\Sigma_s\Sigma_m})
\end{align}
where $tr$ is the trace of a matrix.

Besides using the entire speech and music sets, we also calculate FAD for each emotion to investigate whether the acoustic similarity exhibits different patterns across emotion categories. This analysis is also performed using the representations from each layer.

\section{Experiments}
\subsection{Layerwise Probing of Emotion Recognition Performance}
\label{sec:layerwiseresult}

\subsubsection{Experimental Settings}
For training the SER and MER models, we use cross-entropy loss as the criterion and the \textit{AdamW} optimizer with a learning rate of $1e{-3}$. Due to the variation in the number of epochs required to achieve optimal validation performance using different SSL representations (particularly layer 11), we train for 500 epochs to ensure all classifiers reach their full potential.
% \YC{too many epochs. Which epoch gives the best performance?}

% Two domains, three upstream models, and 12 layers yield 72 runs. The first subsection discusses the implications from all-emotion results and the second that from per-emotion results.

\subsubsection{Results and Discussions}
Fig.~\ref{fig:layerwise} presents the \textbf{all-emotion} results. It can be observed that:

\textbf{\textit{(I)}} For SER, our results align with previous findings, showing that while W2V2 experiences a performance decline in its final layers, HuBERT does not \cite{li2023exploration, lin2023prosody}. In fact, performance improves in the deeper layers of HuBERT compared to its shallow layers. Additionally, we observe that MERT behaves similarly to HuBERT, exhibiting better performance in deeper layers than W2V2. This is likely due to the pretraining objectives and strategies of MERT aligning closely with those of HuBERT. For MER, all models behave consistently with SER. Moreover, the speech models (W2V2 \& HuBERT) perform better at SER, while the music model (MERT) is better at MER, which is reasonable given the differences in their pretraining data.

\begin{figure}[ht]
    \centering
    \includegraphics[width=\columnwidth]{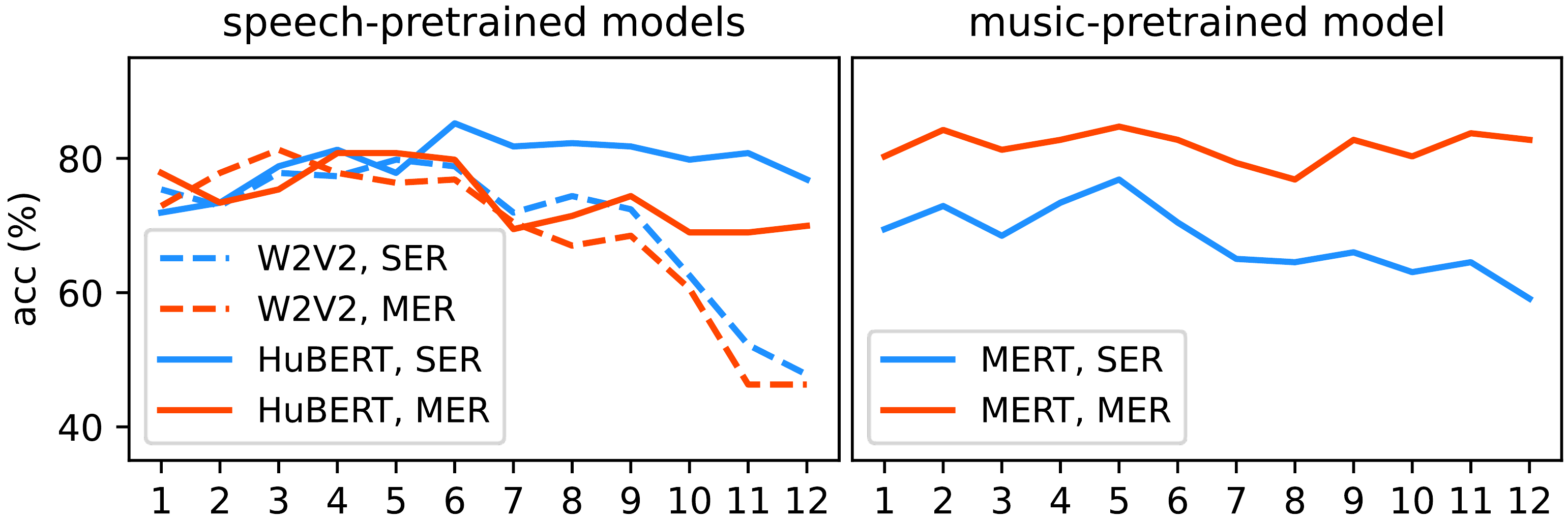}
    \caption{Layerwise SER and MER accuracy using SSL representations.}
    \label{fig:layerwise}
\end{figure}

\textbf{\textit{(II)}} For the speech models, the accuracy gap between SER and MER after the middle layers is more pronounced in HuBERT than in W2V2. Since both models are pretrained on the same data, and the lexical content of RAVDESS does not affect SER or MER, it is likely that HuBERT’s deeper layers retain acoustic information that W2V2 misses yet is emotion-relevant. This retained acoustic information could explain the larger performance gap between HuBERT’s SER and MER performance from the middle layers. Such a gap is not observed in W2V2, even though its SER performance remains better than its MER performance.

\begin{table}[ht]
\caption{Performance summary (UA\%) of SER. \textbf{Bold}: best performance per row. Parentheses: best layer.}
\label{tab:layerwise speech}
\centering
    \begin{tabular}{lccc}
        \toprule
        Models & W2V2 & HuBERT & MERT\\
        \midrule
        Best & 79.80 (5) & \textbf{85.22} (6) & 76.85 (5) \\
            Worst & 47.78 (12) & 71.92 (1) & 59.11 (12)\\
        Mean & 70.28 & \textbf{79.31} & 67.82\\ \bottomrule
    \end{tabular}
    \centering
\end{table}
% \vspace{-7pt}
\begin{table}[ht]
\caption{Performance summary (UA\%) of MER. \textbf{Bold}: best performance per row. Parentheses: best layer.}
\label{tab:layerwise song}
\centering
    \begin{tabular}{lccc}
        \toprule
        Models & W2V2 & HuBERT & MERT\\
        \midrule
        Best & 81.28 (3) & 80.79 (4, 5) & \textbf{84.73} (5)\\
        Worst & 46.31 (11, 12) & 68.97 (7) & 76.85 (8) \\
        Mean & 68.51 & 74.26 & \textbf{81.81}\\ \bottomrule
    \end{tabular}
    \centering
\end{table}

\textbf{\textit{(III)}} MERT shows a significant initial difference between SER and MER, which is not observed on speech models. Despite MERT having similar training objectives and structure to HuBERT, this suggests that the acoustics from the shallow layers of speech models are domain-agnostic and largely shareable, whereas those from the shallow layers of MERT are relatively domain-specific. However, given MERT's fairly good performance on SER, this still indicates shared acoustics between music and speech. Additionally, the drop in SER performance in the deeper layers of MERT suggests that the features in those layers are more favorable for MER than SER. It is possible that the deep layers of MERT encode high-level music-specific cues, such as those related to tonality and rhythm \cite{mizuno2007neural,parncutt2014emotional}. Therefore, speech SSL models, particularly the shallow layers, may be further trainable with music acoustics, whereas further training of music SSL models with speech acoustics may be less effective.

Table \ref{tab:layerwise speech} and \ref{tab:layerwise song} summarize the overall performances of the speech models and music model on both SER and MER.

\begin{figure}[ht]
    \centering
    \includegraphics[width=\columnwidth]{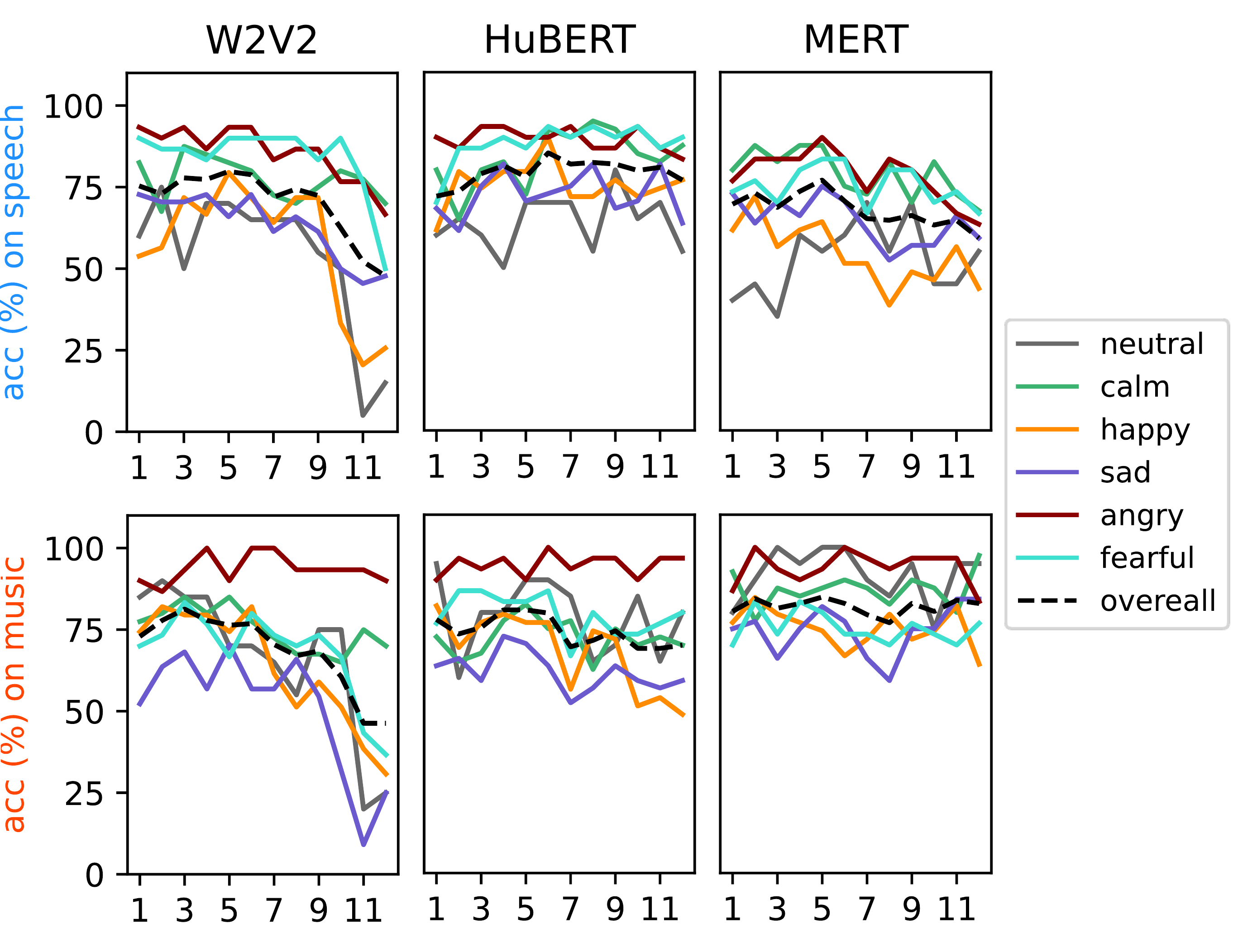}
    \caption{Layerwise SER (top) and MER (bottom) accuracy per emotion using representations from the SSL models.}
    % \YC{1. can you make legend on the right a bit smaller and make more space for the figures on the left? 2. change ``song'' to ``music''.}
    \label{fig:emo acc}
    % \vspace{-5pt}
\end{figure}

Since different emotions exhibit distinct paralinguistic patterns, for instance, \textit{angry} and \textit{happy} typically have higher intensity and pitch, while \textit{sad} and \textit{calm} tend to have lower intensity \cite{lugger2007relevance,li2007stress}, \cite{li2023exploration} demonstrated that W2V2 shows emotion bias by not encoding different emotional speech equally. In light of this, we compare and contrast emotions by calculating their respective recognition accuracies. Fig.~\ref{fig:emo acc} shows the \textbf{per-emotion} results.
% , from which we can see that:

For SER, it can be observed that performance on \textit{angry} and \textit{fearful} speech is generally the best, followed by \textit{calm}, across all speech and music models. This suggests that both speech and music models favor the acoustic characteristics of these emotions in speech. However, for MER, recognizing \textit{fearful} and \textit{calm} proves to be more challenging, particularly in the case of \textit{fearful} with MERT. In contrast, \textit{angry} music becomes easier to recognize, even for speech models. Moreover, \textit{neutral} performs much better on music with MERT than any other \textit{neutral} cases. Additionally, \textit{angry} music with W2V2 does not show a performance drop, unlike other emotions with W2V2 for either speech or music. We believe that the deep layers of W2V2 might be particularly effective at modeling acoustic characteristics related to \textit{angry} music, such as intense rhythm. These observations suggest that different emotions in speech and music exhibit distinct acoustic characteristics, highlighting the emotion bias problem in both speech and music SSL models, and strengthening the findings of \cite{li2023exploration}. Finally, other emotions all have their respective variations on speech or music with speech models or the music model, but the overall patterns remains similar. This further verifies previous findings that shared acoustic representations exist between speech and music \cite{zhang2015recognizing,gomez2020emotionrecognition}.

\begin{table*}[ht]
\centering
\large
\caption{Emotion recognition results (UA\%) with domain adaptation using the two-stage fine-tuning. Left of $\rightarrow$: stage one. Right of $\rightarrow$: stage two. \textbf{Bold}: best performance within respective two-stage fine-tuning process. \textcolor{orange}{\textbf{Orange}}: best performance of the speech models. \textcolor{cyan}{\textbf{Blue}}: best performance of the music model.}
\label{tab:domain}
\scalebox{0.9}{
\begin{tabular}{l|cccc|cccc|cccc}
\hline
\multicolumn{1}{c|}{\multirow{2}{*}{Approach}} & \multicolumn{4}{c|}{W2V2} & \multicolumn{4}{c|}{HuBERT} & \multicolumn{4}{c}{MERT} \\ \cline{2-13} 
\multicolumn{1}{c|}{} & \multicolumn{2}{c|}{\textit{SER $\rightarrow$ MER}} & \multicolumn{2}{c|}{\textit{MER $\rightarrow$ SER}} & \multicolumn{2}{c|}{\textit{SER $\rightarrow$ MER}} & \multicolumn{2}{c|}{\textit{MER $\rightarrow$ SER}} & \multicolumn{2}{c|}{\textit{SER $\rightarrow$ MER}} & \multicolumn{2}{c}{\textit{MER $\rightarrow$ SER}} \\ \hline
Baseline & 73.89 & \multicolumn{1}{c|}{69.95} & 65.02 & 74.88 & 83.25 & \multicolumn{1}{c|}{76.35} & 68.97 & 84.24 & 62.56 & \multicolumn{1}{c|}{83.74} & 85.22 & 61.58 \\ \hdashline
WS & 79.31 & \multicolumn{1}{c|}{77.83} & 73.40 & 80.30 & 80.30 & \multicolumn{1}{c|}{81.28} & 78.82 & 82.76 & 61.85 & \multicolumn{1}{c|}{89.66} & 84.23 & 72.77 \\ \hdashline
PEFT & 87.20 & \multicolumn{1}{c|}{\textbf{87.68}} & 79.80 & \textbf{87.20} & 88.18 & \multicolumn{1}{c|}{\textbf{91.13}} & 91.62 & \textbf{\textcolor{orange}{93.10}} & 85.22 & \multicolumn{1}{c|}{\textbf{\textcolor{cyan}{93.60}}} & \textbf{91.63} & 82.76 \\ \hline
\end{tabular}}
\end{table*}

\subsection{Domain Adaptation for Cross-Domain Performance Improvement}

\subsubsection{Experimental Settings}
For the baseline and WS models, we use the \textit{AdamW} optimizer with a learning rate of $1e{-3}$. For the PEFT models, we use the \textit{AdamW} optimizer with a learning rate of $1e{-4}$. For the same reason as for the layerwise probing, each model is trained for 300 epochs at each stage to achieve its best possible performance.
% \YC{too many epochs. Which epoch gives the best performance?}

\subsubsection{Results and Discussions}

Table~\ref{tab:domain} summarizes the domain adaptation results using all three approaches and models. Note that the results of SER and MER from the same model are not comparable, as they are different tasks. We only compare SER with SER and MER with MER. It can be seen that for the speech models (W2V2 and HuBERT), fine-tuning on either domain in stage one always helps the other domain in stage two. For example, the performances of SER followed by MER (74.88, 80.30, 87.20 for W2V2; 84.24, 82.76, 93.10 for HuBERT) are always better than directly conducting SER (73.89, 79.31, 87.20 for W2V2; 83.25, 80.30, 88.18 for HuBERT). Furthermore, PEFT works extremely effectively for HuBERT, achieving the best performance (93.10) and the largest improvement (4.92). For the music model, while several performances are improved after domain adaptation (e.g., 72.77 vs. 61.85 for SER and 93.60 vs. 91.63), such a phenomenon does not generally hold. This finding further verifies our conclusion of Finding \textbf{\textit{III}} in Section~\ref{sec:layerwiseresult} that \textit{speech SSL models are further trainable with music acoustics, yet further training of music SSL models with speech acoustics may be less effective}.

\subsection{Layerwise Analysis of Cross-Domain Acoustic Similarity}
\subsubsection{Experimental Settings}
As FAD is a reference-free measurement, there is no model training involved. We separate the speech and music sets per emotion, and use the \href{https://github.com/microsoft/fadtk}{FAD toolkit}
% FAD toolkit\href{https://github.com/microsoft/fadtk}{https://github.com/microsoft/fadtk}
for the implementation. The lower the FAD score, the more similar the two sets of representations.

\subsubsection{Results and Discussions}
From Fig.~\ref{fig:fad}, it can be observed that:

\textbf{\textit{(I)}} In terms of distance ranking, all the speech and music models exhibit consistent patterns: \textit{angry} and \textit{fearful} have the smallest distances, followed by \textit{happy}, \textit{sad}, \textit{neutral}, and \textit{calm}. These consistent patterns indicate that the SSL models have a similar ability to capture acoustic similarities related to emotion in their respective representations. Since speech and music share the same text in RAVDESS, the low FAD scores for certain emotions indicate that their acoustics are largely similar between speech and music. For example, \textit{angry} speech and music are more acoustically similar than \textit{calm} speech and music.

\textbf{\textit{(II)}} HuBERT and MERT exhibit aligned patterns not only in the per-emotion ranking but also in the overall trend. The FAD score increases from the first layer to the last, which can be attributed to the similar structures of HuBERT and MERT. However, several middle layers show divergent patterns: while the score continues to rise in HuBERT, it remains stable in MERT. This difference is related to their respective capabilities: HuBERT encodes phonetic-level information in the middle layers \cite{pasad2023comparative}, whereas such capability has not been found in MERT. Additionally, the FAD scores for the final layers of both HuBERT and MERT rise sharply, indicating that their ability to capture shared acoustic features between speech and music weakens at these layers.

\textbf{\textit{(III)}} Unlike HuBERT and MERT, the FAD scores for W2V2 exhibit a significant shift starting from the middle layers. One explanation for this is that W2V2 begins encoding word identity and meaning between the middle and final layers, as noted by \cite{pasad2021layer}. Since speech and music datasets share identical text, the acoustic similarity increases, resulting in lower FAD scores. Additionally, different emotions behave differently in layer 11, where \textit{calm}, \textit{neutral}, and \textit{sad} emotions show a sharp rise, while \textit{angry}, \textit{happy}, and \textit{fearful} do not. It appears that layer 11 amplifies these acoustic differences, consistent with previous findings that observed peculiar behavior in this layer due to its contrastive masked segment prediction function \cite{pasad2021layer,li2023exploration}. We further demonstrate the peculiarity of layer 11 in that it is particularly sensitive to cross-domain acoustic differences.

Since using FAD does not involve model training, thereby avoiding potential impacts on the fairness of comparison across emotions, the results and findings instead further underscore the emotion-bias problem present in SSL representations.

\begin{figure}[htb]
    \centering
    \includegraphics[width=\columnwidth]{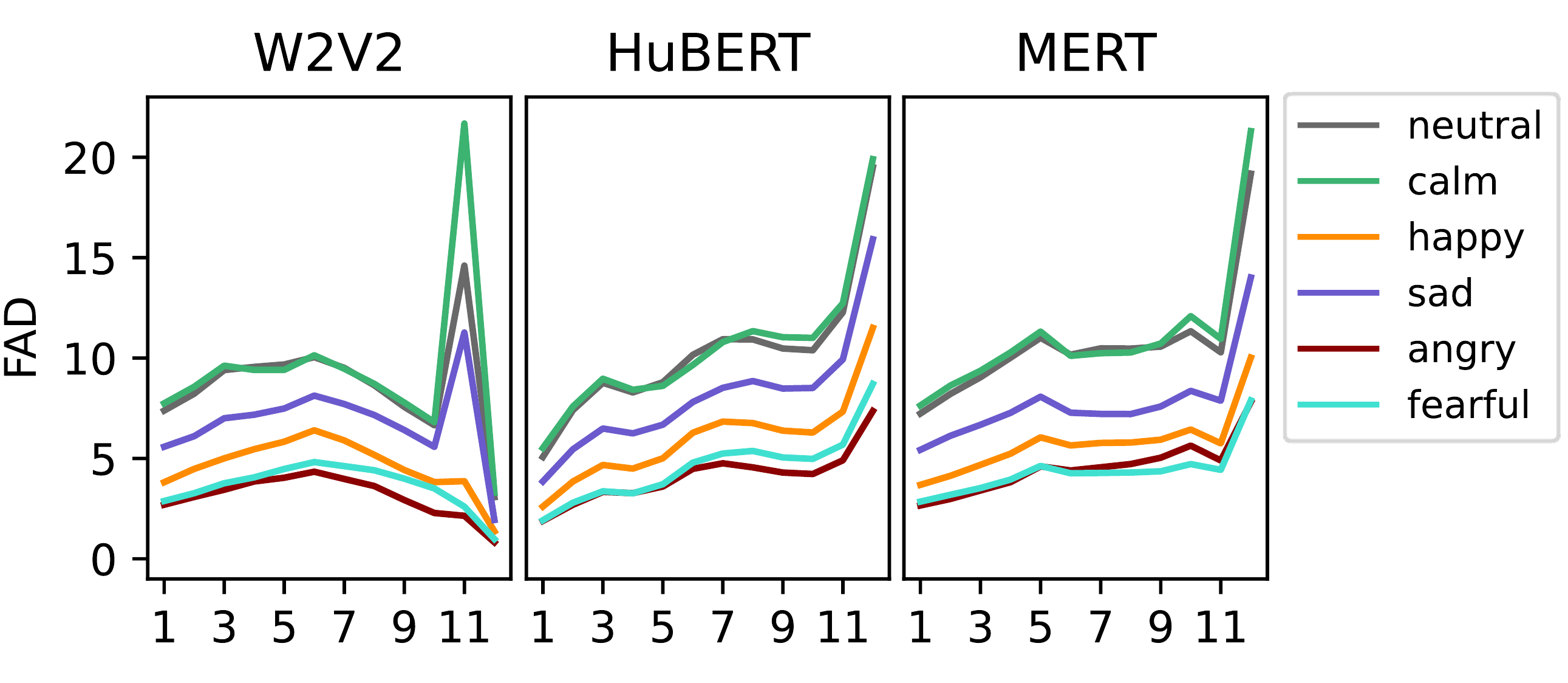}
    % \vspace{-10pt}
    \caption{Layerwise cross-domain FAD per emotion of the SSL models.}
    % \YC{can you make legend on the right a bit smaller and make more space for the figures on the left?}
    \label{fig:fad}
    % \vspace{-5pt}
\end{figure}

\section{Conclusion}
As the first work exploring acoustic similarity between emotional speech and music through SSL models, we conduct layerwise probing to explore the models' ability to capture emotion-relevant acoustic cues and uncover the commonality and differences between speech and music SSL models. Moreover, aiming to leverage each domain to enhance the emotion recognition performance for the other domain, we implement and compare three approaches, demonstrating efficacy of the PEFT approach for cross-domain adaptation and noticing the domain-specificity of the models. Finally, we employ FAD to measure the acoustic similarity between speech and music representations on each emotion, revealing different behaviors of SSL models and the emotion-bias issue in both domains. Our study provides new insights into the acoustic similarity between emotional speech and music, particularly with SSL models. We also demonstrate the potential to leverage knowledge from each domain to facilitate the other. As there are limited datasets suitable for further investigation and fair comparison, we acknowledge the limitations of our work and plan to collect new datasets for extending our study in the future.

\balance
\bibliographystyle{IEEEbib}
\bibliography{refs}

\end{document}